\documentstyle[aps,prl,floats]{revtex} 

\begin{document}
\draft
\preprint{}
\twocolumn[\hsize\textwidth\columnwidth\hsize\csname@twocolumnfalse%
\endcsname

\title{The Spectrum of the two dimensional
 Hubbard model at low filling}

\author{Henrik Bruus\cite{HB} and
Jean-Christian Angl\`es d'Auriac\cite{JCAdA}}
\address{Centre de Recherches sur les Tr\`es Basses Temp\'eratures,
CNRS, B.P.\ 166, F-38042 Grenoble C\'edex 9, France}
\date{July 5, 1995}

\maketitle

\begin{abstract}
Using group theoretical and numerical methods we have calculated
the exact energy spectrum of the two-dimensional
Hubbard model on square lattices with four electrons for
a wide range of the interaction strength.
All known symmetries, i.e.\ the full space group symmetry,
the SU(2) spin symmetry, and, in case of a bipartite lattice, the SU(2)
pseudospin symmetry, have been taken explicitly into account.
But, quite remarkably, a large amount of residual degeneracies
remains  giving strong evidence for the existence of a yet
unknown symmetry. The level spacing distribution and the spectral
rigidity are found to be in close to but not exact agreement with
random matrix theory. In contrast, the level velocity correlation
function presents an unexpected exponential decay qualitatively
different from random matrix behavior.
\end{abstract}

]

The surprising discovery of high temperature
superconductivity in complicated cuprates containing planes of
conducting electrons has renewed the interest in the study of
two dimensional strongly interacting electronic systems.
One of the simplest models describing such systems
is the Hubbard model \cite{Hubbard}, but in spite of its
apparent simplicity this model turns out to be extremely difficult to
fully understand (see e.g.\ the  recent review by E.\
Dagotto \cite{Dagotto}).  This is mainly due to the
lack of a small parameter which makes the use of well established
perturbative methods in condensed matter theory highly
questionable. In this state of affairs the importance of performing
numerical calculations of the spectrum for finite clusters has grown.

On the other hand the concept of level statistics has proven
itself to be a useful tool in the understanding of non--perturbative
many-body quantum systems. Originally applied to models in nuclear
physics \cite{Mehta} the method consists of describing {\em some\/}
spectral properties of the Hamiltonian by those of a well suited ensemble
of random matrices. The ensemble of random matrices to be chosen
depends on the physical system under consideration. For example
a system which can be solved by the Bethe Ansatz displays the same
distribution, $P(s)$, of the energy level spacings $s$ as an
ensemble of random diagonal matrices, i.e.\ a Poisson distribution
$P(s)=e^{-s}$. In contrast $P(s)$ for non-integrable
time reversal symmetric systems has been found empirically to be the
same as that for the Gaussian Orthogonal Ensemble (GOE), i.e.\ a
Wigner distribution $P(s) = (s\pi/2) \exp(-s^2\pi /4)$.
The appearance of a Wigner distribution is
interpreted in terms of level repulsion: states belonging to the
same symmetry class repel each other. This gives
a connection between quantum chaos and the level distribution.

In this letter we extend the level statistical analysis of quantum
Hamiltonians \cite{Montambaux,Hsu,Faas} to include the two dimensional
Hubbard model with four electrons (low filling) as a function of the
coupling strength $U/t$ to be defined below.
Our study is not restricted only to comprise $P(s)$; the
spectral rigidity $\Delta_3(\lambda)$ is also investigated in detail
as well as the level velocity correlation $c(x)$, a quantity related
to the deformation of the spectrum when $x \sim U/t$ is varied. A
careful group theoretical analysis enables us to sort the states with
respect to {\em all\/} known quantum numbers and makes possible the
numerical diagonalization of the
Hubbard Hamiltonian on lattices as big as the $6\!\times\!6$
square lattice. Residual degeneracies in the resulting
spectra constitute the first numerical evidence of the existence of a
new  unknown symmetry of the Hubbard model. Further traces of this
new symmetry are seen as small deviations from the expected random matrix
behavior of $P(s)$ and $\Delta_3(\lambda)$.

{\em Group theoretical and numerical analysis.}
Throughout this letter we study
the one--band Hubbard model containing
nearest--neighbor hopping and on--site interaction:
\begin{equation} \label{HubbardModel}
\hat{H} = -t \hat{T} + U \hat{V} \equiv -t \sum_{\langle
i,j \rangle,\sigma} \hat{c}^{\dagger}_{j\sigma} \hat{c}_{i\sigma} + U \sum_{i}
\hat{n}_{i\!\uparrow}  \hat{n}_{i\!\downarrow}.
\end{equation}
We treat the case of a two dimensional $L\!\times\!L$ square lattice
with periodic boundary conditions. To
investigate the low filling properties of the model we
restrict ourself to four electrons.
We vary $L$ from 3 to 6 and obtain
the filling factors 0.22, 0.13, 0.08, and 0.06 respectively.
To achieve the
a priori maximal reduction of the problem we construct the symmetry
projection operators
corresponding to all known symmetries of the model and use them to
project into symmetry invariant subspaces of the full Hilbert
space \cite{Tinkham}. For our study it is essential to keep
all symmetries in the model, rather than adding
extra terms to $\hat{H}$ and sorting out the symmetries after
diagonalization as is commonly done. This makes
the group theoretical analysis more complex, but it leads to
larger reductions, and it yields more precise numerical results. To
facilitate further the calculation of spectra for arbitrary values of
$U/t$ we calculate and store matrix elements of the operators
$\hat{T}$ and $\hat{V}$ rather than of $\hat{H}$. Then for any given
value of $U/t$ the spectrum $E(U/t)$ as well as the derivative
$\partial E/\partial(U/t)$ is calculated by straightforward
diagonalization of $-t \hat{T} + U \hat{V}$ \cite{BruusDauriac}.

The first symmetry we consider is the space group $G_L$ of the lattice.
It consists of all permutations $g$
mapping any neighboring sites $i$ and $j$ onto
neighboring sites $g(i)$ and $g(j)$.
To each element $g$ of $G_L$ an operator $\hat{g}$
in the Hilbert space
can be associated in a straightforward manner forming
a group  $\hat{G}_L$ of operators. For any lattice size $G_L$ has been
analyzed in detail
by Fano, Ortolani, and Parola \cite{Fano}
and found to be $D_L \!\otimes\! D_L
\makebox[1.0em]{\small $\bigcirc$}\hspace{-1.0em}\makebox[1.0em]{s}
Z_2$, where $D_L$ is the usual dihedral group of index $L$.
However, this result is {\em not\/} valid for the special
case $L=4$, where the spatial group is three times larger.
In this work we use the correct $G_4$ \cite{BruusDauriac}. To deal
with the space symmetry we employ the
projection operators $\hat{\cal P}^{(R)}_k$ of row $k$ in representation
$R$ of $G_L$ having the usual form $\frac{l_R}{h}\sum_g
\Gamma^{(R)*}_{kk}(g) \: \hat{g}\;$
\cite{Tinkham}.

Next is the SU(2) spin symmetry.
Since $\hat{H}$ commutes with the total spin $\hat{\bf{S}}$
and with the corresponding raising and lowering operators $\hat{S}_+$ and
$\hat{S}_-$, we work in the $\hat{\bf{S}}_z = 0$
sector, i.e.\ with two up spins and two down spins.
Moreover, $\hat{\bf{S}}$ commutes with all operators of $\hat{G}_L$, and
the combination of the two groups is a direct product. The spin
symmetry  of four--electron states is dealt with through the
projections $\hat{\cal P}^{(S)} |a\!\uparrow, b\!\uparrow, c\!\downarrow,
d\!\downarrow \rangle$ having the form $\sum_{\pi}
\alpha^{\pi}_{abcd} |\pi_a\!\uparrow, \pi_b\!\uparrow, \pi_c\!\downarrow,
\pi_c\!\downarrow \rangle$, where $\pi$ is a permutation
of the sites $abcd$ \cite{BruusDauriac}.

The last of the known symmetries is the SU(2)
pseudospin symmetry. This symmetry of dynamical
origin based  on the $\eta$--paring
mechanism was discovered recently \cite{CNYang}.
It exists only for bipartite lattices, which for periodic square
lattices demands $L$ to be even.
The generators of the SU(2) pseudospin symmetry are:

\begin{equation} \label{Jdef}
\hat{J}_- =\sum_i (-1)^i
\hat{c}_{i\!\uparrow} \hat{c}_{i\!\downarrow}, \;
\hat{J}_+ = \hat{J}_-^{\dagger},\;
\hat{J}_z = \frac{1}{2} (\hat{N} - L^2),
\end{equation}

\noindent
where $\hat{N}$ is the electron number operator.
The pseudospin  $\hat{J}$ commutes
with $\hat{H}$ as well as with all
$\hat{g} \in \hat{G}_L$ and $\hat{S}$.
For  $\hat{J}$ we find the projection
$\hat{\cal P}^{(J)}
|a\!\uparrow,
b\!\uparrow, c\!\downarrow, d\!\downarrow \rangle$
to be of the form $\sum_{\pi}
\beta^{\pi}_{abcd} |\pi_a\!\uparrow, \pi_b\!\uparrow, \pi_c\!\downarrow,
\pi_c\!\downarrow \rangle$, where
$\pi_a \pi_b \pi_c \pi_d$ are sites related to $abcd$ by the
pair hopping operator $\hat{J}_+ \hat{J}_- + \hat{J}_z^2 - \hat{J}_z$
\cite{BruusDauriac}.

A detailed analysis shows that combining the spin and the pseudospin
symmetries yields a SO(4) symmetry rather than a SU(2) $\otimes$ SU(2)
symmetry \cite{YangZhang}, however, the projection operators
still form direct products.
The full symmetry group for even $L$ is ${\cal G} =
G_L\otimes{\rm SO(4)}$, and in addition to the principal energy
quantum number $n$ the states are labeled with the three quantum
numbers $R$, $S$, and $J$ corresponding to the total projection
operator $\hat{\cal P}^{(R)}_k \otimes \hat{\cal P}^{(S)} \otimes
\hat{\cal P}^{(J)}$. For $L$ odd ${\cal G} = G_L\otimes{\rm SU(2)}$,
and only $R$ and $S$ are defined.
In row 2 and 3 of Table \ref{SizeAndDeg} we show the dimension of
the total Hilbert space and the much smaller dimension of the largest
symmetry invariant subspace found by the group theoretical analysis.

\begin{table}[htb]
\begin{tabular}{|ll||r|r|r|r|}
$L$ & & 3 & 4 & 5 & 6 \\ \hline \hline
\multicolumn{2}{|l||}{Dim($\cal H$)}
& 1,296 & 14,000 & 90,000 & 396,600 \\ \hline
\multicolumn{2}{|l||}{Dim(${\cal I}$)}
& 38 & 146 & 1,794 & 5,490 \\ \hline \hline
$S$=0,1& deg.\ $U$-indep. & 0 & 3,662 & 3,139 & $^*$13,120 \\
$S$=0,1& deg.\ $U$-dep. & 0 & 38 & 0 & $^*$19 \\ \hline
$S$=0,1& non-deg.\ $U$-indep & 9 & 62 & 572 & $^*$64 \\
$S$=0,1& non-deg.\ $U$-dep. & 1,161 & 8,818 & 73,639 & $^*$50,418
\\ \hline
$S$=2  & $U$--indep. & 126 & 1,820 & 12,650 & 58,905 \\
\end{tabular}
\vspace{5pt}
\caption{\label{SizeAndDeg}
For  $L=3,4,5$, and 6 are shown the dimension
Dim($\cal H$) of the total unreduced Hilbert space, the dimension
Dim(${\cal I}$) of the largest symmetry invariant subspace, and the
number of degenerated and non--degenerated levels dependent or
independent of $U/t$ for the three values of the total spin $S$ {\em
after\/} taken all known symmetries into account. The numbers marked by
'*' in the $L=6$ column covers only 36 out of 130 representations
(18\% of the states).
}
\end{table}

{\em A new symmetry.}
The most remarkable fact revealed by Table \ref{SizeAndDeg} is the
large amount of residual degeneracies present {\em after\/} sorting the
spectrum according to all known symmetries.
We  note that states with the maximal spin $S=2$ contain no
doubly occupied sites. They are thus {\em independent\/} of $U$ and
are consequently excluded from our study. States with $S=1$ ($S=0$)
can accommodate up to one (two) doubly occupied site(s) and in
general they are therefore expected to depend on $U/t$. Nevertheless,
it turns out that
there exist states with $S=0$ and $S=1$ which are independent of
$U$. This is the first evidence that an additional symmetry exists in
the problem. Much stronger evidence comes from the existence of
degeneracies remaining  after excluding accidental
inter--representational degeneracies as well as the trivial $l_R$--fold
degeneracy of the $l_R$ rows in a given  representation of
the space group.
This result is summarized in row 4 to 7 of
Table~\ref{SizeAndDeg} showing the total number of $U$--(in)dependent
and (non--)degenerated states.
It is seen that almost all the residual degenerated states are
$U$--independent. In fact only for even $L$ we found a few $U$--dependent
degenerated states, and they all belong to invariant subspaces with a
dimension smaller than 7.
We stress that finding
the residual degeneracies was only possible because of the
relatively small
size of the symmetry invariant subspaces enabling us to employ the
standard iterative diagonalization techniques \cite{BruusDauriac}. It
is extremely difficult to find degeneracies using the
Lanczos method.

{\em Level statistics.}
Based on Table \ref{SizeAndDeg} we conjecture that the new symmetry
is related to the $U$--independent states. We therefore exclude all
such states from our treatment. The results we show in the figures
below are obtained with $L=5$ because this is the largest lattice for
which all states have been computed. The results for other values of
$L$ are similar.

The first step in the level statistical analysis is the `unfolding' of
the spectrum in order to transform the energies $E_n$
into `reduced energies' $\varepsilon_n$ of constant
density. This amounts to carefully computing the average cumulative
density of states $N_{\rm av}(E)$ from the actual cumulative
density of states \cite{Mehta,BruusDauriac}.
To study the statistical properties of the spectrum at small energy
scales we calculate $P(s)$.
The result for three different values of
$U/t$ spanning six orders of magnitude is
shown in Fig.~\ref{PdS}. In all three cases $P(s)$
is close to the Wigner distribution; it possesses a pronounced linear
level repulsion for small $s$, a peak near $s=1$ signaling spectral
rigidity, and a Gaussian tail. However, the
statistics are good enough to note a significant discrepancy, and in
fact on a 99.99\% confidence level a $\chi^2$ test leads to a rejection
of the hypothesis that $P(s)$ is Wigner distributed for any of the
three values of $U/t$. The level repulsion
is not as strong as expected from a GOE spectrum.
We interpret this as another trace of the remaining
symmetry, and we speculate that if we could sort the states according
to the new symmetry, then the level repulsion would be stronger and
$P(s)$ would be closer to the Wigner distribution. It should be noted,
though, that the new symmetry does not mix the symmetry classes very
much, so in a sense the known symmetries nearly sort the spectrum
perfectly, and in fact for some individual representations for $L=6$
a $\chi^2$ test does not lead to rejection \cite{BruusDauriac}.

\begin{figure}[htb]
\vspace*{65mm} 
\caption{\label{PdS}
The probability distribution $P(s)$ of the level spacings $s$ in the
unfolded $5\!\times\!5$ Hubbard model spectrum averaged over all
symmetry sectors in the three cases of small,
medium, and large interaction strength $U/t$.
The full line is the Wigner distribution found for
GOE random matrices.}
\end{figure}

To study the statistical properties of the spectrum  on larger energy
scales we have computed the Wigner-Dyson level rigidity
$\Delta_3(\lambda)$ defined as the least square deviation of the level
staircase $N_{\rm av}(\varepsilon)$ from the best fitting straight
line in an interval of length $\lambda$ \cite{Mehta}:
\begin{equation} \label{Delta3def}
\Delta_3(\lambda) = \left\langle
\frac{1}\lambda \min_{(A,B)}
\int_{\varepsilon_0-\frac{\lambda}{2}}^{\varepsilon_0+\frac{\lambda}{2}}
(N_{\rm av}(\varepsilon)-A\varepsilon-B)^2d\varepsilon
\right\rangle_{\varepsilon_0} \! \! \!.
\end{equation}
The brackets denote an averaging over $\varepsilon_0$.
Fig.~\ref{fig_rig} shows $\Delta_3(\lambda)$
of one specific symmetry sector for five different
values of the interaction strength $U/t$ ranging from 1, close to the
integrable non--interacting case, to 20, well into the strongly
interacting regime. When $\lambda$
is small the rigidity of the Hubbard spectrum is very close to that of
the GOE random matrices, while for larger $\lambda$ the spectra
becomes less rigid;
i.e.\ on a large energy scale the levels become
uncorrelated, even though adjacent levels are strongly correlated.
The reason for this behavior is that the strong degeneracies existing at
$U/t=0$ is slowly lifted as $U/t$ is increased. For small values of
$U/t$ levels within each degeneracy band interact. Hence for small
$\lambda$ only properly randomized levels are sampled by
$\Delta_3(\lambda)$. If $\lambda$ is larger than the typical width of
the bands, $\Delta_3(\lambda)$ samples also the non--random gaps. As
$U/t$ increases the gaps close in and for larger and larger $\lambda$
only randomized levels are sampled by $\Delta_3(\lambda)$.
The spectral rigidity can thus be used
to monitor this regular to random transition in the spectrum. The value
$\lambda^{*}(U/t)$ below which the rigidity has the GOE form grows
roughly linear with $U/t$.

\begin{figure}[htb]
\vspace*{80mm} 
\caption{\label{fig_rig}
The level rigidity $\Delta_3(\lambda)$ of the ($R$=6,$S$=0)--sector
of the $5\!\times\!5$ Hubbard model for five
different values of $U/t$ compared to the random diagonal matrix
ensemble (dashed straight line) and to the GOE (full line). For each
data set are shown two typical error bars.
The error bars grow as a function of increasing
$\lambda$ and as a function of decreasing $U/t$.
}
\end{figure}

{\em The level velocity correlation.}
To obtain a more quantitative measure of the $U/t$--dependence of the
spectra, we calculate the level velocity function $c(x)$ \cite{Szafer}.
The generalized conductance
$C(0) = \langle ( \frac{\partial \varepsilon_i}{\partial U/t}
)^2 \rangle_{i,U/t} $ and the rescaled interaction
strength $x = \sqrt{C(0)} U/t$ are introduced and $c(x)$ is defined as:
\begin{equation} \label{cofx}
c(x) \equiv
\left\langle
\frac{\partial \varepsilon_i(\tilde{x}+x)}{\partial \tilde{x}}
\frac{\partial \varepsilon_i(\tilde{x})}{\partial \tilde{x}}
\right\rangle_{i,\tilde{x}}.
\end{equation}
It has been found that $c(x)$ is universal for a variety of
single--particle systems with many different external perturbations
given that the systems exhibit random matrix
behavior \cite{Szafer,Simons,BruusStone}. Even for some
many--body systems the
same result has been obtained,  and it has been suggested that the
spectra of all nonintegrable strongly correlated systems can be
classified according to the generalized conductance, mean--level
spacing, and Dyson ensemble \cite{Faas}.
However, we find that this is not the case for the Hubbard model.
In Fig.~\ref{plotcofx} $c(x)$ is shown for the representation
($R$=6,$S$=0) with $L=5$. This particular representation was chosen
since out of its 630 states only 2 were independent of $U$.
It is seen that $c(x)$ is an exponential
decaying function as opposed to the expected generic GOE-curve.
In our model the random matrix behavior stems from choosing a
sufficiently large value of $U/t$, e.g.\ $U/t = 20$ as seen by the
results of $P(s)$ and $\Delta_3(\lambda)$. The parametric change originates
from a further change in $U/t$. We have chosen to do the analysis in
the interval $U/t \in [20,50]$. We can conclude that random matrix
behavior of $P(s)$ and $\Delta_3(\lambda)$ is not a sufficient criterion for
observing universal behavior in the level velocity correlation
function for many--body systems.

\begin{figure}[htb]
\vspace*{50mm} 
\caption{\label{plotcofx}
The level velocity correlation function $c(x)$ (dots) of the
($R$=6,$S$=0)--sector of the $5\!\times\!5$ Hubbard model compared to the GOE
result (dashed line). The data are obtained with $U/t \in [20,50]$ for
level 350 to 400.  Five typical error bars are shown. The insert shows
the same data on a semi--logarithmic scale revealing an exponential
decay of $c(x)$ for $x>0.2$.
}
\end{figure}

{\em Conclusion and discussion.}
The existence of a new symmetry in the Hubbard model at low filling
has been demonstrated numerically by projecting the Hamiltonian into
invariant subspaces of all the known symmetries for a wide range of
$U/t$ and noting how a significant amount of degeneracies persists. The
commonly used statistical analysis of random matrix spectra has been
applied. A small discrepancy between the actual $P(s)$ and
$\Delta_3(\lambda)$ the expected random matrix results have been
interpreted as a consequence of the new symmetry. We have demonstrated
that the Hubbard model is a specific example of a many--body model
with strong interaction where, eventhough $P(s)$ and
$\Delta_3(\lambda)$ are close to the GOE behavior, the
parametric dependence $c(x)$ of the spectra is qualitatively different
from random matrix systems. We speculate that GOE--like behavior would
be obtained by taking the new symmetry properly into account and, in
the case of $c(x)$, by adding a next--nearest neighbor interaction
compatible with the symmetry group to introduce a more rapid
and stronger mixing  between the levels.

It is important to establish the exact nature of the new
symmetry, and to find out if it exists for higher filling factors,
where the physical properties of the model are known to be different.
In particular it would be interesting to study the regime near half
filling and to extend our analysis to more specific Hamiltonians like
the $tJ$--model and the Heisenberg model. The method presented here is
general and can be applied to these models widely used in studies of
high temperature superconductivity and magnetism in two dimensions.

We want to thank Benoit Dou\c{c}ot for stimulating discussions.
H.B.\ was supported by the European Commission under grant no.\
ERBCHBGCT 930511.




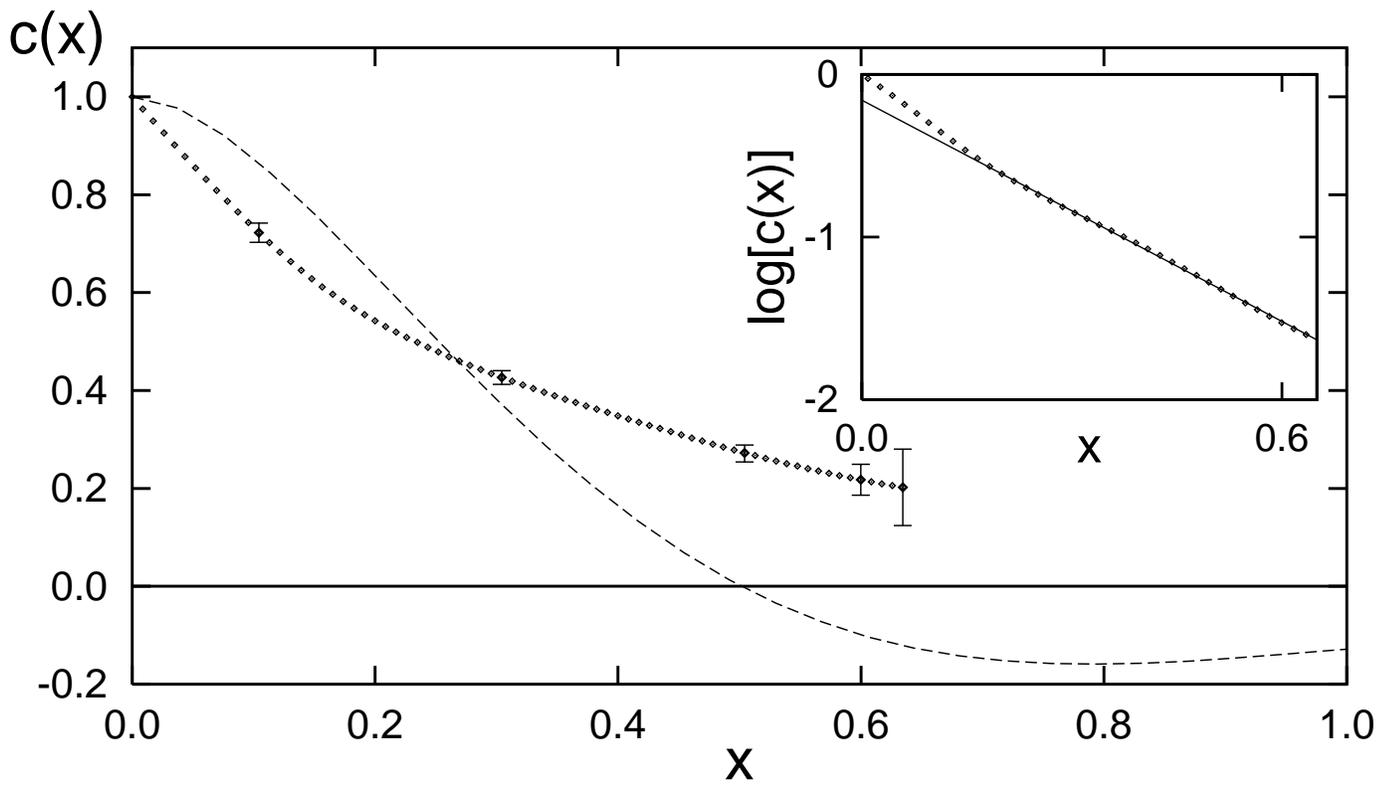
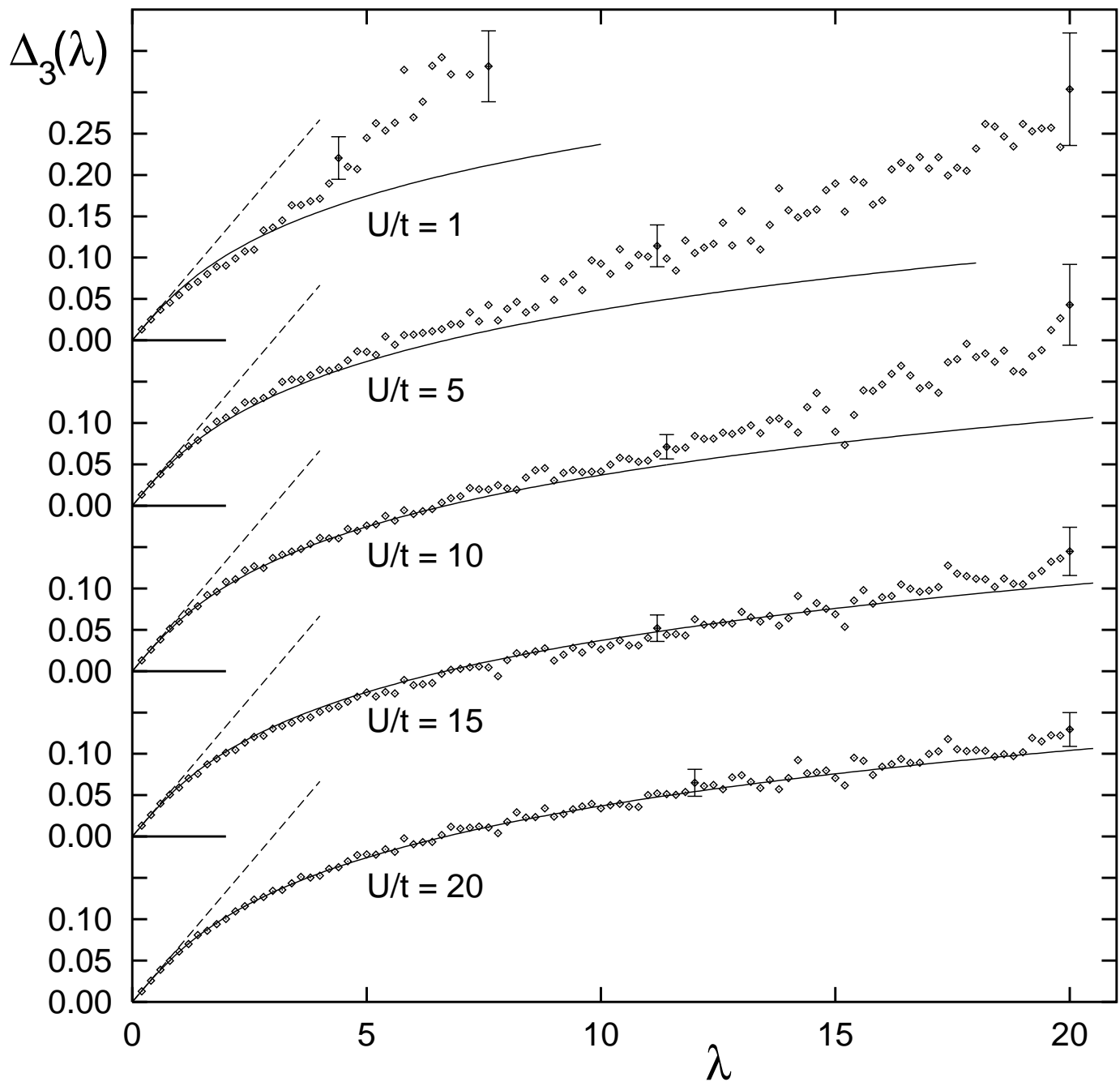
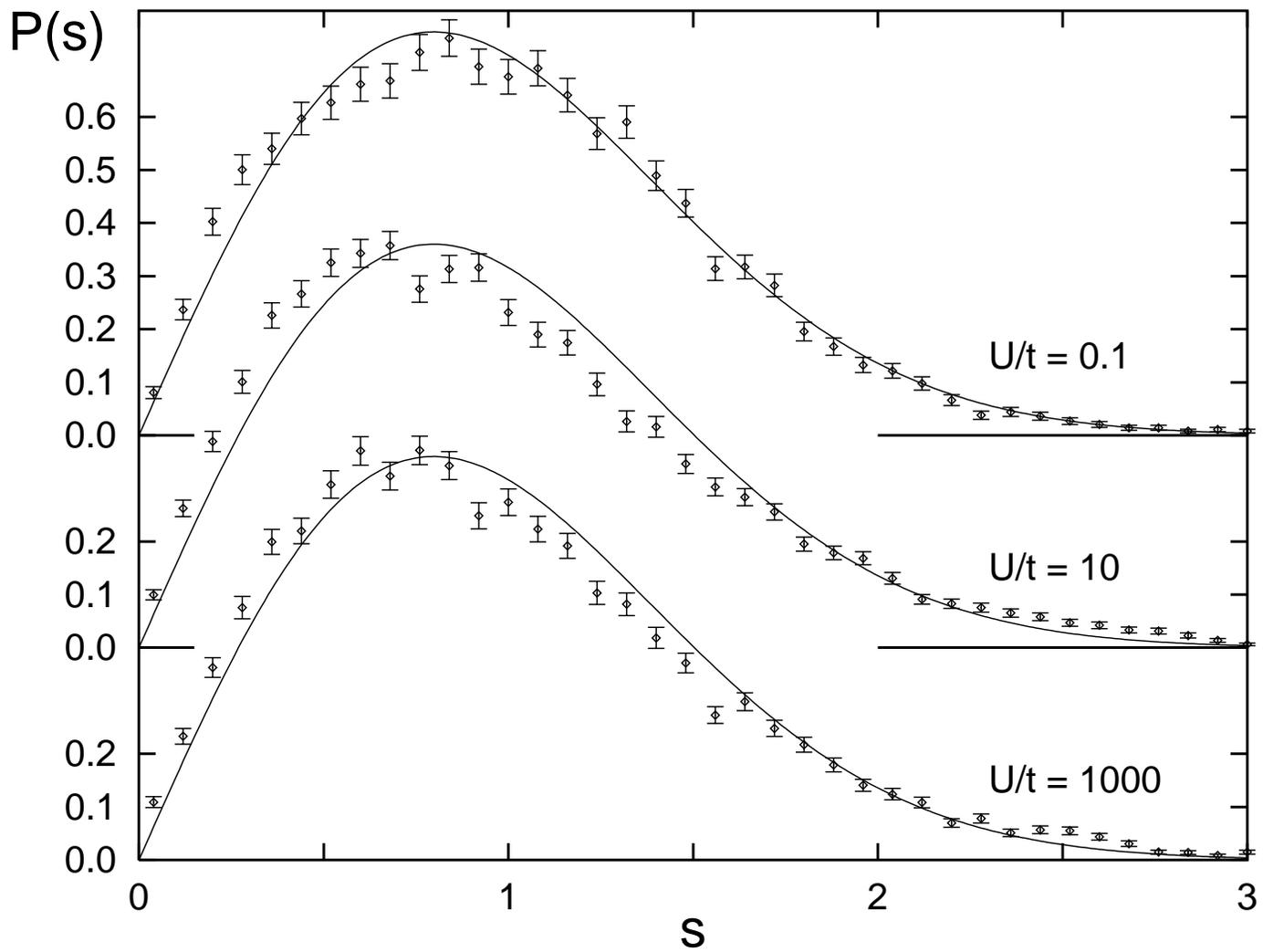
\end{document}